\documentclass[conference]{IEEEtran}

\def\BibTeX{{\rm B\kern-.05em{\sc i\kern-.025em b}\kern-.08em
    T\kern-.1667em\lower.7ex\hbox{E}\kern-.125emX}}
    
\usepackage[]{hyperref}
\usepackage{caption}
\usepackage{subcaption}
\usepackage{multirow}
\usepackage{array}
\usepackage{algorithm}
\usepackage{algorithmicx}
\usepackage{algpseudocode}
\usepackage{graphicx}
\usepackage{amsmath} 
\usepackage{booktabs}

\newcommand{\sysname}{Hecaton}
\newcommand{\sota}{state-of-the-art}

\begin{document}

\title{\sysname: Training Large Language Models with Scalable Waferscale Chiplet Systems}

\author{\IEEEauthorblockN{Zongle Huang}
\IEEEauthorblockA{\textit{Dept. of Electronic Engineering} \\
\textit{Tsinghua University}}
\and
\IEEEauthorblockN{Shupei Fan}
\IEEEauthorblockA{\textit{Dept. of Electronic Engineering} \\
\textit{Tsinghua University}}
\and
\IEEEauthorblockN{Chen Tang}
\IEEEauthorblockA{\textit{Dept. of Electronic Engineering} \\
\textit{Tsinghua University}}
\and
\IEEEauthorblockN{Xinyuan Lin}
\IEEEauthorblockA{\textit{Dept. of Electronic Engineering} \\
\textit{Tsinghua University}}
\and
\IEEEauthorblockN{Shuwen Deng}
\IEEEauthorblockA{\textit{Dept. of Electronic Engineering} \\
\textit{Tsinghua University}}
\and
\IEEEauthorblockN{Yongpan Liu}
\IEEEauthorblockA{\textit{Dept. of Electronic Engineering} \\
\textit{Tsinghua University}}
}

\maketitle

\begin{abstract}
Large Language Models (LLMs) have achieved remarkable success in various fields, but their training and finetuning require massive computation and memory, necessitating parallelism which introduces heavy communication overheads. Driven by advances in packaging, the chiplet architecture emerges as a potential solution, as it can integrate computing power, as well as utilize on-package links with better signal integrity, higher bandwidth, and lower energy consumption. However, most existing chiplet-related works focus on DNN inference. Directly porting them to LLM training introduces significantly large quantities of DRAM access and network-on-package (NoP) overheads which make state-of-the-art chiplet designs fail, highlighting a research gap.

This work proposes \sysname, a scalable and cost-effective chiplet system for LLM training. We first provide a chiplet architecture with tailored scheduling that can largely reduce DRAM accesses. We further design an efficient distributed training method that reduces NoP communication complexity and relieves constraints on SRAM capacity and layout. Theoretical analysis shows that the entire system achieves \textit{weak scaling}: as the workload and hardware resources grow proportionally, the computation-to-communication ratio remains nearly constant. Experiments with various workloads and hardware configurations verify the property, and \sysname{} achieves $5.29\times$ performance improvement and $3.46\times$ energy reduction on Llama3.1-405B, compared to the tensor parallelism in Megatron. To the best of our knowledge, we propose the first chiplet architecture specifically used for LLM training or finetuning, with guaranteed performance regardless of the problem scale.

\end{abstract}

\pagestyle{plain} 

\renewcommand{\thefootnote}{}
\footnotetext{{Correspond to: Zongle Huang \textless huangzl23@mails.tsinghua.edu.cn\textgreater}}
\renewcommand{\thefootnote}{\arabic{footnote}}

\section{Introduction}

Large Language Models (LLMs) have achieved significant success across a wide range of applications in recent years~\cite{vaswani2017attention,dosovitskiy2020image,devlin2018bert,brown2020language} but pose challenges in computing power and memory capacity. The scaling law of LLMs~\cite{kaplan2020scaling} demonstrates that models with more parameters exhibit better performance, which leads to higher hardware requirements. Training an LLM needs significantly large datasets, immense computing power, and substantial memory that stores intermediate activations, weights, their gradients, and optimizer states. For instance, training the Llama model with 1.4T tokens takes 2048 A100 GPUs 34 days, creating terabyte-scale memory usage~\cite{touvron2023llama}. Finetuning, on the other hand, is to train a pre-trained model on a smaller, task-specific dataset, and has a more extensive demand. Although finetuning converges faster, its run-time memory usage and dataflow remain almost unchanged, which still makes it challenging to deploy.

\begin{figure}[h]
  \centering
  \includegraphics[width=\linewidth]{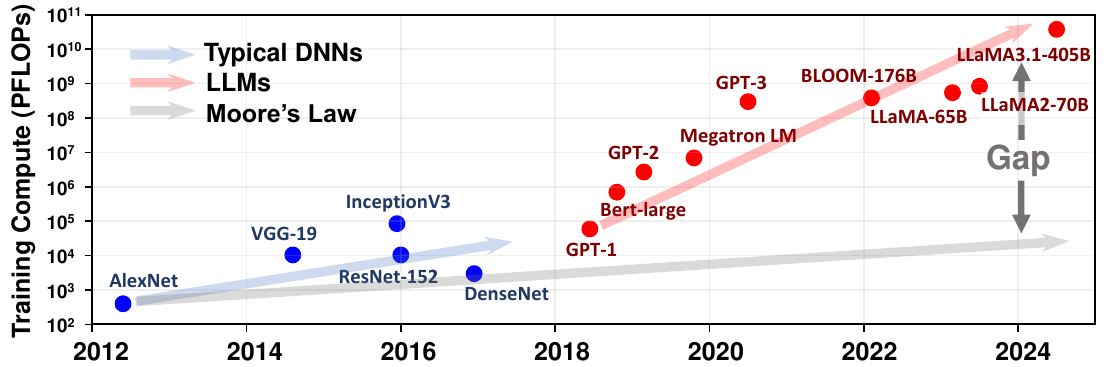}
  \caption{The growing speed of training FLOPs and hardware}    
  \label{fig:model-scaling}
\end{figure}


Chiplet architecture offers an approach for integrating a large number of computing resources, but its performance could be hurt by communication overheads. In chiplet architecture, multiple 
small dies are manufactured separately and then integrated within the same package~\cite{zimmer20190,huang2021wafer,mahajan2016embedded}, which helps mitigate the hardware restriction imposed by the reticle limit~\cite{naffziger2021pioneering}, thus achieving scaled-out computing power. The overall die-to-die~(D2D) connections form a network-on-package~(NoP). Compared to off-package connections such as NVLink~\cite{nvlink} or InfiniBand~\cite{infiniband}, on-package D2D links have the potential to achieve better signal integrity\footnote{Signal integrity~(SI) measures the quality of an electrical signal. More details can refer to~\cite{shin2023signal}.}, higher density, and lower energy consumption due to their shorter communication distances and more stable channels~\cite{sharma2022universal}. 
However, in chiplet architecture, multiple dies process the same workload in a distributed manner, therefore introducing significant communication overheads. For example, measurements on the Simba~\cite{shao2019simba} silicon prototype showed that when 36 dies are integrated, the NoP overhead accounts for over 50\%, severely affecting the system's scalability.

Existing works attempt to mitigate communication overheads of chiplets but fail to deal with massive DRAM accesses and NoP communication. 
These works either improve task scheduling through exhaustive design space exploration~\cite{tan2021nn,cai2024gemini,cai2024abss,hao2023monad}, or optimize NoP topology to reduce energy consumption and transmission latency~\cite{laskar2024enhancing,feng2023heterogeneous,feng2023scalable,yin2023aries}.
However, these works primarily target traditional deep neural networks~(DNNs) or inference, which differs significantly from LLM training and cannot be directly ported because: 
First, the dataflow and dependencies in training or finetuning are more complex than those in inference which introduces lots of DRAM accesses that are not considered by \sota{} works.
Second, the size of LLMs (billion to trillion) vastly exceeds that of DNNs (most under 100 million), thus not only requiring larger on-chip buffers but also new training method to tackle NoP communications. Additionally, the paradigm of LLMs and DNNs have root differences where the Transformer architecture introduces the novel multi-head attention mechanism not used by DNN.

In this paper, we introduce \sysname\footnote{
Hecatoncheires, with their hundred hands, symbolize immense power and multitasking capabilities in Greek mythology, aligning well with the concept of a scalable and efficient chiplet architecture.}, 
a scalable and cost-effective chiplet system targeting LLM training and finetuning with high utilization. 

To address the DRAM-access and paradigm challenges, we first provide a chiplet template with a tailored scheduling and adopt distributed buffers that save weights collaboratively and carefully design the dataflow. 
The hardware is scalable, as we utilize adjacent connections instead of package-size interposer for D2D links, which is hard to manufacture with huge package size. We also modify the NoP router to achieve a higher transfer throughput. The architecture is cost-effective, as it only uses DDR5 DRAM surrounding the package instead of expensive high bandwidth memories (HBMs)~\cite{kim2014hbm}. To compensate its lower bandwidth, we exploit layer fusion~\cite{alwani2016fused} and on-and-off-package overlap to orchestrate computation and DRAM access. By decoupling software tasks from hardware execution units, \sysname{} enable training with arbitrarily large batch sizes.


We further propose a novel distributed training method to reduce NoP communication overheads and relieve constraints on SRAM capacity or layout for chiplet. 
Through the co-design of 2D matrix tiling and communication scheme, we reduce the amount of data that needs to be transferred and achieve high utilization of D2D links. 
Compared to the tensor parallelism used in existing works such as Megatron~\cite{shoeybi2019megatron} or Optimus~\cite{song2023optimus}, this method reduces asymptotic communication complexity.

We provide a theoretical analysis that \sysname{} exhibits weak scaling, which refers to that, as the model and hardware resources scale proportionally, the main components of system latency\textemdash computation, NoP communication, and memory access\textemdash maintain nearly constant proportion. The reduced communication complexity of our method allows the NoP overheads to scale proportionally with other system components as the problem size increases.
Our method also ensures that the required SRAM capacity per die remains unchanged. 
This alleviates concerns that the distributed system may be bottlenecked by communication overheads, or that the currently used dies become inadequate when dealing with extremely large models in the future.

We evaluate the whole design on workloads with different scales including Bert-Large~\cite{devlin2018bert}, Bloom-1.7B\cite{le2023bloom}, GPT3-6.7B\cite{brown2020language} and Llama2-70B~\cite{touvron2023llama2}. The simulator supports various hardware configurations, including different numbers and layouts of dies, packages, buffer sizees and computation resources. The functional units are realized in RTL, while the attributes of D2D links are sourced from Universal Chiplet Interconnect Express (UCIe) standards~\cite{UCIe}. 

The main contributions of our work are as follows:

\begin{itemize}
    \item Propose a scalable and cost-effective chiplet architecture for LLM training and finetuning. Combined with scheduling optimizations, \sysname{} can train with arbitrarily large batch sizes. To the best of our knowledge, this is the first work systematically discussing how to train LLMs with chiplet. (Section~\ref{chap:overview})
    \item Design an efficient distributed training method that reduces asymptotic communication complexity and relieves constraints on SRAM capacity and layout compared to existing methods. (Section~\ref{chap:method})
    \item Provide a theoretical proof that the entire system exhibits weak scaling. The property means that \sysname{}'s performance is guaranteed regardless of the model size. (Section~\ref{chap:theo})
    \item Evaluate \sysname{}'s performance and observe the weak scaling predicted by the theory. 
    Compared to the tensor parallelism used in Megatron, \sysname{} achieves 5.29× throughput and 3.46× energy efficiency improvements in Llama2-70B. (Section~\ref{chap:eval})
\end{itemize}

\section{Background}

\subsection{Chiplet Architectures}

As the size of monolithic chips approaches the reticle limit and the yield of advanced process nodes declines~\cite{naffziger2021pioneering}, chiplet architectures have emerged as a promising approach to provide scaled-out computation power~\cite{naffziger20202}. They integrate multiple smaller dies to construct a large system, offering reduced costs, simplified design, and easier verification. 

\begin{figure}[h]
  \centering
  \includegraphics[width=0.85\linewidth]{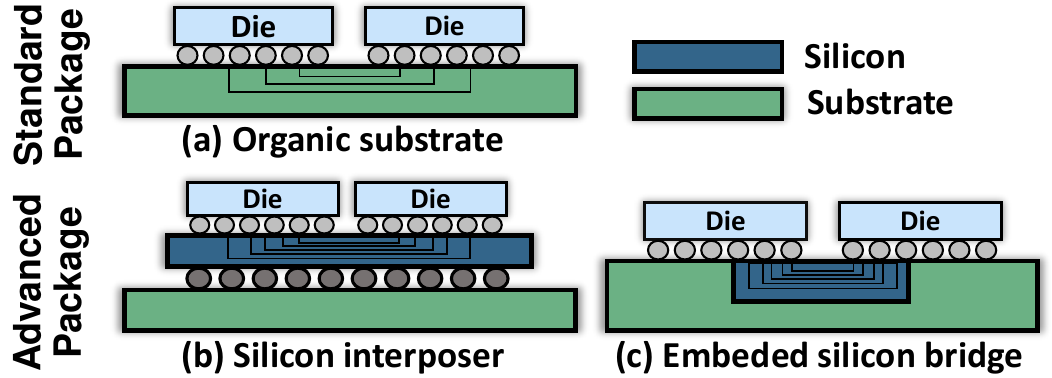}
  \caption{Typical packages of chiplets.}
  \label{fig:packaging}
\end{figure}

The core of chiplet architecture lies in packaging technologies that facilitate D2D communication, with typical forms shown in Figure~\ref{fig:packaging}. Standard packaging connects dies through traces on organic substrates~\cite{zimmer20190,lall1993overview}, which is cost-effective but results in relatively low bandwidth. By introducing silicon as the fabric for routing, advanced packaging can achieve higher interconnect density and lower power consumption. Two representative approaches are silicon interposers~\cite{huang2021wafer} and embedded silicon bridges~\cite{mahajan2016embedded}. The interposer-based packaing introduces a silicon layer with area comparable to the entire chiplet, allowing for flexible routing between dies in different locations. However, when the number of dies becomes very large, its manufacturing cost increases significantly, making it less scalable for larger systems. On the other hand, embedded silicon bridges connect only adjacent dies, resulting in lower costs and better scalability. However, this approach sacrifices interconnect flexibility, as long-distance transfers require multiple hops to be realized. The bandwidth of each D2D link is computed as a multiplication of transfer rate and interface width.

\subsection{Transformer Workload}
\label{chap:transformer}

\begin{figure}
    \centering
    \includegraphics[width=\linewidth]{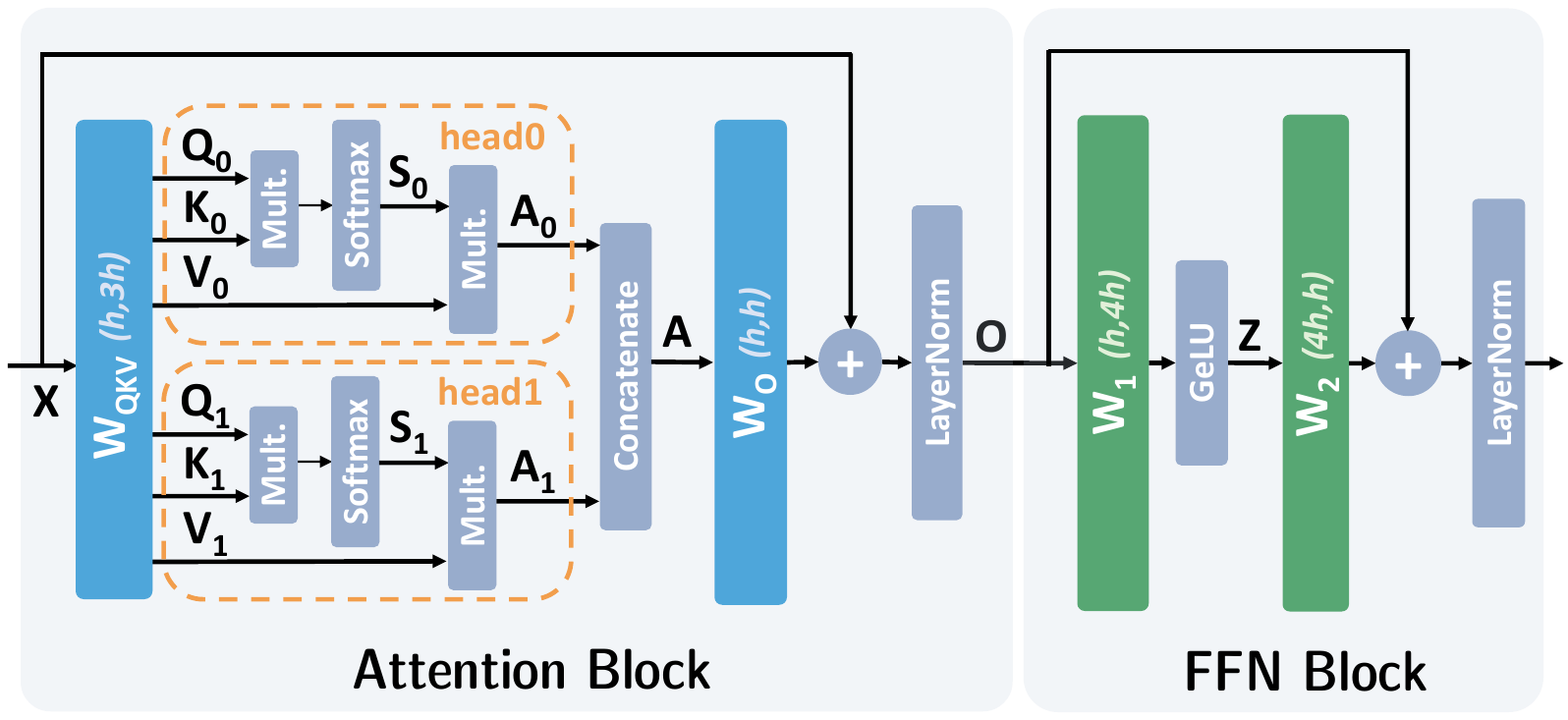}
    \caption{A Transformer layer containing an Attention block FFN block. Names of activations and weights are annotated.}
    \label{fig:Transformer}
\end{figure}

LLMs are composed of several Transformer layers stacked together, each containing an Attention block and a Feed-Forward Network~(FFN) block as shown in Figure~\ref{fig:Transformer}. Most activations are tensors with three dimensions: batch size~($b$), sequence length~($s$), and hidden size~($s$). During training, both $b$ and $s$ are adjustable configurations, while $h$ is determined by the model and reflects the Transformer's representational capacity.

In Attention blocks, input $X$ multiplies with weight matrix $W_{QKV}$ to generate three types of activations: query ($Q$), key ($K$), and value ($V$). These activations are divided into multiple segments along the hidden size, forming different heads. Each head computes a scaled dot-product attention $S_i = \textit{\text{softmax}}(Q_iK_i^T/s)$ and $A_i=S_iV_i$, where subscript $i$ indicates the head index and $s$ is a scaling factor. $A_i$ from all heads are concatenated to form $A$, which then undergoes a linear transformation with weight $W_O$ and adds $X$ passed by the residual link. Layer normalization (LayerNorm) is performed to improve training stability. Notably, unlike linear layers with trainable weights, the operands in the attention mechanism are dynamically generated during each forward pass.

FFN blocks consist of an up-scaling and down-scaling linear layer, with a non-linear function like GeLU inserted between them. The intermediate activation $Z$'s hidden size is often scaled four times. Each FFN block also includes a residual connection and a normalization layer.

\subsection{Parallel LLM Training and Finetuning}
\label{chap:parallelism}

Parallelism is essential for LLM training and finetuning to reduce computation time and accommodate massive data, which can be futher categorized into data parallelism (DP), pipeline parallelism (PP), and tensor parallelism (TP).

In DP, each device stores a copy of model parameters and processes part of the batch~\cite{li2014scaling}. The weight gradients computed by each device are aggregated to update weights. It speeds up computation without significantly reducing memory usage. PP~\cite{huang2019gpipe,narayanan2019pipedream} distributes LLM layers across different devices, effectively reducing memory usage per GPU. However, pipeline bubbles form at the beginning and end of the computation, which hurts hardware utilization.

\begin{figure}[!t]
  \centering
  \includegraphics[width=\linewidth]{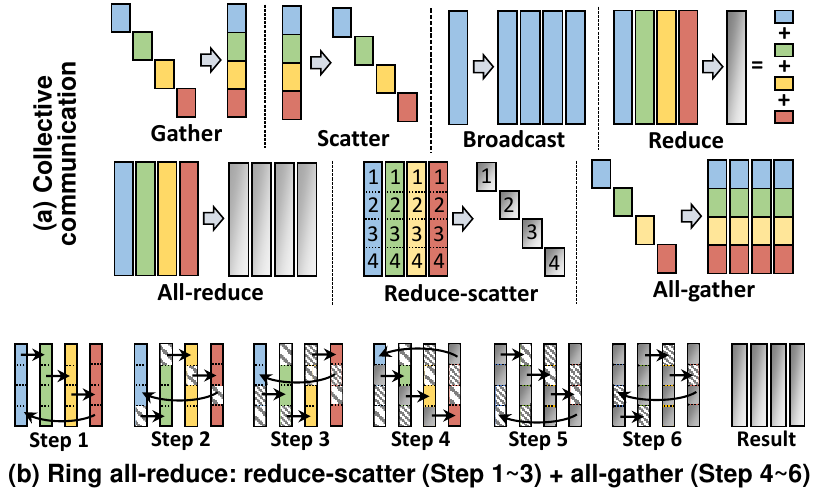}
  \caption{Collective communication and the ring all-reduce.}
  \label{fig:all-reduce}
\end{figure}

TP divides weight matrices and assigns them to different devices. Collective communications (CC) are necessary to aggregate the results computed by each machine. Typical CCs are demonstrated in Figure~\ref{fig:all-reduce}(a). Depending on the partitioning dimension, TP can be further categorized into 1D-TP such as Megatron~\cite{shoeybi2019megatron} and 2D-TP such as Optimus~\cite{xu2023efficient}. The CCs they require are listed in Table~\ref{tab:sum}.

We display the operations of ring all-reduce algorithm~\cite{baidu} in Figure~\ref{fig:all-reduce}(b). This algorithm splits all-reduce into reduce-scatter and all-gather stages and can achieve full bandwidth utilization. Assume the total data volume is $S$ and there are $N$ machines, then in each step, each machine transfers a data chunk with the size of $\frac{S}{N}$ to its adjacent in altogether $2(N-1)$ steps. Therefore, transmission time $T$ has:
\begin{equation}
    T = \text{\small $\frac{S}{N\times \text{bandwidth}}\times2(N-1) \propto  \frac{N-1}{N}$}
    \label{equ:ring}
\end{equation}

There are other algorithms performing all-reduce. 2D-torus ring executes simultaneous vertical and horizontal ring all-reduce on a 2D-torus topology~\cite{tanakaimagenet,ying2018image}. Hybrid ring executes grouped and hierarchical transfers, more suitable for layers with less parameters such as CNN~\cite{jia2018highly}. These works are also summarized in Table~\ref{tab:sum}.

\begin{table}[h]
\caption{Summary of TPs and corresponding algorithms}
\label{tab:sum}
\small
\begin{tabular}{p{1.6cm}p{1.8cm}p{1.4cm}p{1.7cm}}
\hline
\centering\arraybackslash \textbf{Tiling dimension} & \centering\arraybackslash \textbf{Representive work} & \centering\arraybackslash \textbf{Required CC} & \centering\arraybackslash \textbf{Algorithm for CC} \\
\hline
 &  &  & \multicolumn{1}{c}{flat-ring~\cite{baidu}} \\
\cline{4-4}
 \centering 1D &  \centering Megatron~\cite{shoeybi2019megatron} &  all-reduce &  \multicolumn{1}{c}{2D-torus~\cite{tanakaimagenet,ying2018image}} \\
\cline{4-4}
 &  &  &  \multicolumn{1}{c}{hybrid-ring~\cite{jia2018highly}} \\
\hline
\multicolumn{1}{c}{\multirow{2}{*}{2D}} & \multicolumn{1}{c}{\multirow{2}{*}{Optimus~\cite{xu2023efficient}}} &  \multicolumn{1}{c}{broadcast,} &  \multicolumn{1}{c}{recursive} \\
 &  &  \multicolumn{1}{c}{reduce} & \multicolumn{1}{c}{doubling} \\
\hline
\end{tabular}
\end{table}

\section{System Overview}
\label{chap:overview}

Section~\ref{chap:arch} introduces the \sysname{} architecture, highlighting its flexibility, scalability, and cost-effectiveness. In order to reduce its corresponding high DRAM access overheads, we optimize its scheduling as demonstrated in Section~\ref{chap:scheduling}.

\subsection{\sysname{} Architecture}
\label{chap:arch}

\begin{figure}
  \centering
  \includegraphics[width=\linewidth]{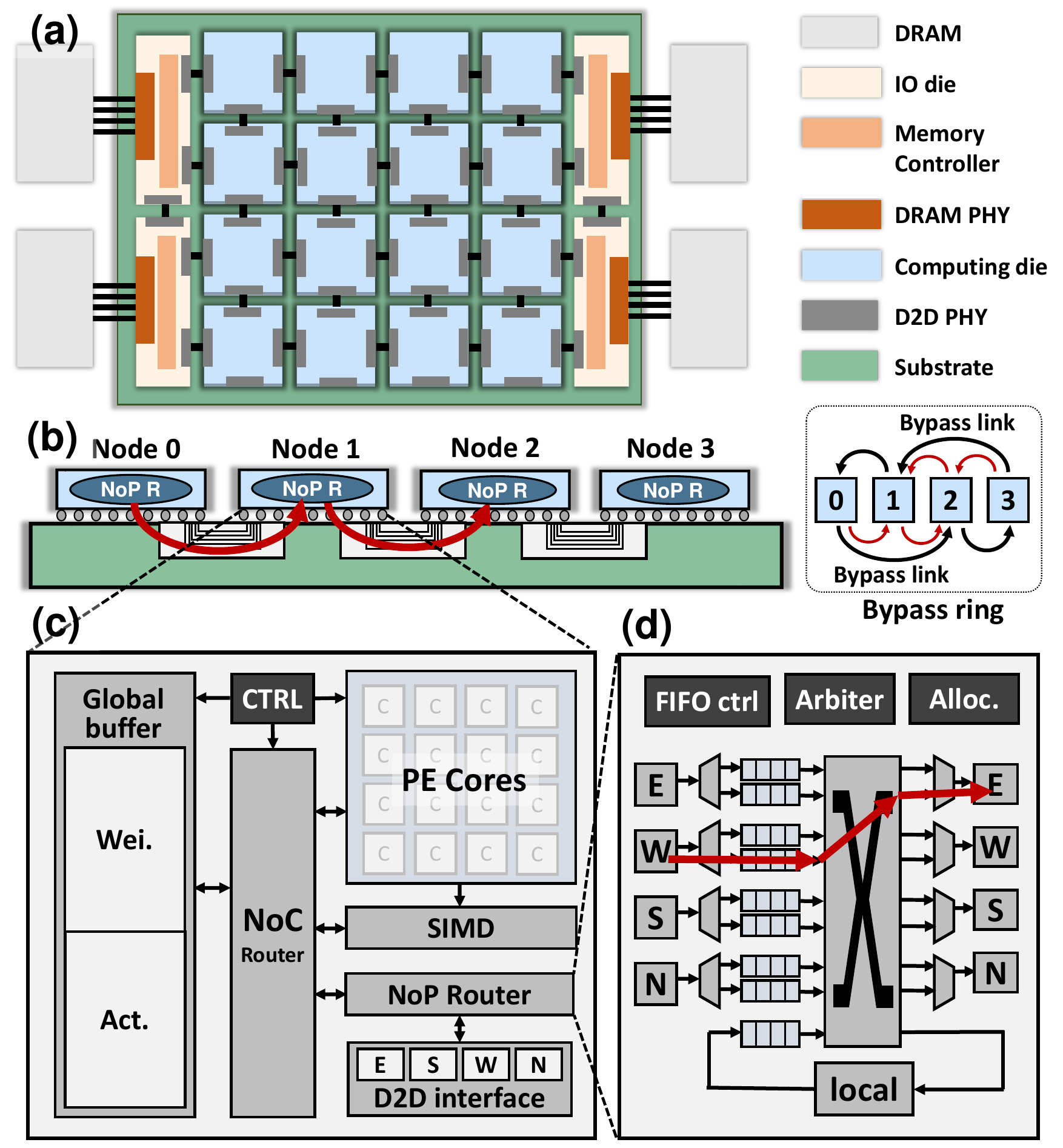}
  \caption{(a)~The architecture of \sysname{} and its main components. (b)~The side view of D2D connection and NoP routers. The red array shows the forwarding from Node~0 to Node~2 via Node~1. We also display the connection fo the proposed bypass ring. (c)~The architecture of computing dies. (d)~The high-throughput NoP router and its bypass channel.}
  \label{fig:architecture}
\end{figure}

Figure~\ref{fig:architecture} shows a typical system, which comprises the following three main components.

\paragraph{Flexible on-package computation.}
Figure~\ref{fig:architecture}(c) shows the architecture of the computing die we use in \sysname. The chiplet design methodology enables a flexible replacement of the computing die with various ASIC accelerators.

Each computing die should consist of the following modules: global weight and activation buffers, a PE array and vector unit for main computation, a NoP router with a corresponding interface, and a controller and network-on-chip (NoC) router for managing intra-die dataflow. The global weight buffers on all dies form a unified on-chip memory pool, together storing parameters of one or more LLM layers. 
The global buffers in \sysname{} are scaled up from those typically used in DNN inference to satisfy the high memory requirements of LLM training, occupying nearly half of the computing die area. 
This work adopts the classic Simba-like~\cite{shao2019simba} structure, but replacing INT8 multiply-accumulators (MAC) with the FP32 version to support high-precision LLM training. The array can also be replaced with variants supporting sparsity and other dataflows~\cite{chen2016eyeriss, yuan2018sticker,song20197,lee2018unpu}, depending on the application. The D2D interface receives or transmits data in different directions and connects to the NoP router, which is introduced later. The NoC router manages communication among PEs and serves as a local interface to the NoP router.

\paragraph{Scalable on-package communication.}
\label{chap:link}
D2D connections and NoP routers within computing dies facilitate the on-package communication. For scalability considerations, we use embedded silicon bridges or organic substrates rather than a complete silicon interposer. As model size grows, the chiplet architecture integrates more dies to provide scaled-out computation power, leading to an expanded package area and high manufacture costs for interposer.

We design low-latency bypass links and high-throughput NoP routers that provide setup for scalable communication. Our distributed method~(Section~\ref{chap:method}) requires a ring topology for dies in each column or row. We propose to implement the ring using bypass links as shown in Figure~\ref{fig:architecture}(b). Compared to the conventional 2D-torus which directly connects dies at two ends, the bypass ring reduces the longest-link latency from the side length to 2 times the adjacent links. 

The router's architecture is shown in Figure~\ref{fig:architecture}(d). It has five ends: local, east ($E$), south ($S$), west ($W$), and north ($N$), managing transmission from various directions. Received packets are buffered in FIFOs and then allocated to a crossbar performing data exchange, overseen by arbitrators and controllers. In the bypass ring, for example, Die~1 not only performs its own data transmission, but also needs to forward data from Die~0 to Die~2. To optimize throughput, we aim to design the router on Die~1 to process these two transactions simultaneously. Observing that the forwarding direction is deterministic\textemdash the receive port is always opposite to the transmit port~($W$ to $E$ or $N$ to $S$)\textemdash only wires connecting specific ports and multiplexers for control need to be added, while other modules are reused.

\begin{figure*}[!t]
    \centering
    \includegraphics[width=\textwidth]{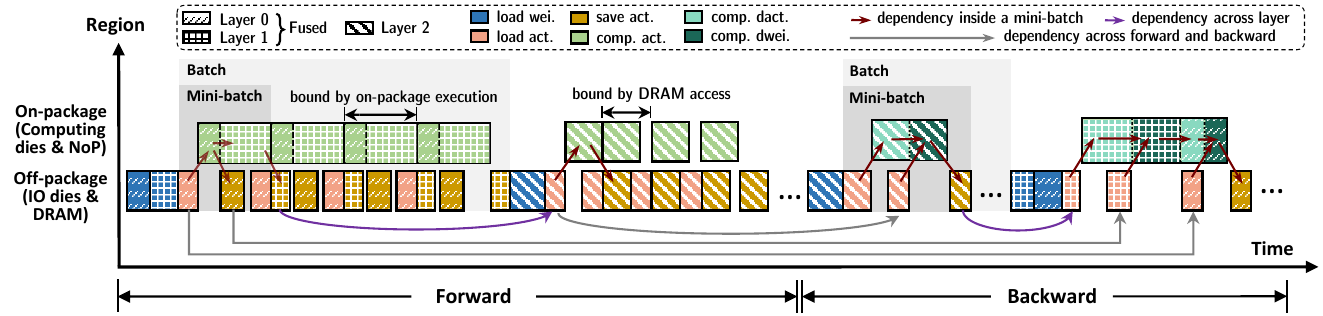}
    \caption{The overall scheduling of \sysname{}, illustrating the training of three layers with a batch divided into four mini-batches. Distinct patterns and colors denote different layers and operations, respectively. Arrays shows various data dependency involved in the training.
    Due to space constraints, we only show operations of one mini-batch in the backward process and omit repetitive parts. In real cases, the latency of backward process should be roughly twice that of the forward process, as it needs to compute gradients of both activation and weight.
    To reduce DRAM overheads, we utilize (1) on-package execution (including computation and communication) and off-package memory access overlap, and (2) layer fusion. 
    }
    \label{fig:scheduling}
\end{figure*}

\paragraph{Cost-effective off-package communication.} 
\label{chap:arch_dram}

We employ cost-effective DDR5 DRAMs instead of expensive HBMs as the system's memory. Under the same budget, DRAMs offer larger storage capacity, which is crucial for LLMs that require extensive memory resources. However, this advantage comes at the expense of reduced bandwidth. To mitigate potential memory access bottlenecks, we employ scheduling techniques which will be demonstrated in Section~\ref{chap:scheduling}. The DRAM accesses are managed by IO dies, which contain memory controllers that oversee transactions from different channels. The system's overall DRAM bandwidth is determined by the product of the number of channels and the bandwidth per channel, with the former being proportional to the package perimeter, and the latter determined by the DRAM type.

\subsection{\sysname{} Scheduling}
\label{chap:scheduling}

We propose the scheduling of \sysname{}
as shown in Figure~\ref{fig:scheduling}. Within a training iteration including the forward and backward process of a batch, we divide a batch into multiple mini-batches as minimal execution units. 
This can help adapt the fixed hardware to arbitrary batch sizes. The larger the activation buffer capacity, the more samples a mini-batch has.
The batch size, which infers the number of samples, is specified by the software training strategy; generally, larger batch sizes lead to more stable training.

We allow weights to be reused by multiple mini-batches, so their DRAM access overhead is amortized and accounts for a small portion of the system. However, the DRAM access for activation is still frequent since it is required for each mini-batch. Unlike inference that only needs to save final results, in training the intermediate activations are also saved, as the backward process uses them to compute the gradient of weights. We use the following two techniques to alleviate the impact of DRAM access for activation on system latency.

\paragraph{On-package execution and off-package memory access overlap.} 
We optimize the throughput by dividing on-package computation / communication and off-package memory access into different pipeline stages, thereby hiding part of the DRAM access latency. The system's critical path is determined by the longer stage. For example, the fused layer in Figure~\ref{fig:scheduling} is bounded by on-packaging execution, while Layer~2 is bounded by off-chip DRAM access. The batch size in training and finetuning is generally kept large to ensure stability, which benefits the pipeline scheduling. 

\paragraph{Layer fusion.}

With layer fusion~\cite{alwani2016fused}, outputs of the current layer is directly used as inputs for the next layer, without the process of saving and loading intermediates from DRAM. For example, in Figure~\ref{fig:scheduling}, Layer~0 and Layer~1 are fused, while Layer~2 is computed separately. A deeper fusion, meaning more layers are executed consecutively, results in a greater reduction in DRAM access. However, the fusion depth is constrained by the capacity of weight buffers. Since the weights of all fused layers need to be stored on-chip, larger weight buffers allow more layers to be fused.

In Transformers, when the weight buffer capacity is tight, all matrix multiplications within the attention blocks are fused, while the two linear layers in the FFN are processed sequentially. This is because the parameter volume of a complete attention block is equivalent to that of a scaling-up or scaling-down FFN layer, both equaling $4h^2$ as illustrated in Figure~\ref{fig:Transformer}. When the weight buffer capacity is sufficient, Attention blocks and FFN blocks can be fused together to further reduce the amount of transferred activation.

\section{Distributed Training Method}
\label{chap:method}
In this section, we propose a distributed training method to perform computation and NoP communication on \sysname{} architecture, i.e., refining the \textit{on-package execution} part in Figure~\ref{fig:scheduling}. It can be viewed as a novel tensor parallelism.

\subsection{Overview}

\begin{algorithm}[!t]
    \caption{One training iteration for a linear layer}
    \label{alg:main}
    {\fontsize{9pt}{12pt}\selectfont
    \begin{algorithmic}
        \State \textbf{Software}: $Y=X\times W, dX=dY\times W^T, dW=X^T\times dY$
        \State \hspace{15mm} $N$ for number of mini-batch
        \State \textbf{Hardware}: $B_A$ for activation buffer, $B_W$ for weight buffer
        
        \State 
        \State \textbf{\textit{// Forward Process}}
        \State scatter $W[j,i]$ to $B_W[i,j]$  \Comment{Step~1}
        \For{$n = 0 \to N-1$}
            \State scatter $X_n[i,j]$ to $B_A[i,j]$ \Comment{Step~2}
            \State all-gather $X_n[i,j]$ within column  \Comment{Step~3}
            \State $\tilde{Y}_n[:,j,i] = X_n[:,j] \times W[j,i]$
            \State reduce-scatter $\tilde{Y}_n[:,j,i]$ within row  \Comment{Step~4}
            \State gather $Y[j,i]$ saved on $B_A[i,j]$  \Comment{Step~5}
        \EndFor

        \State 
        \State \textbf{\textit{// Backward Process}}
        \State scatter $W[i,j]$ to $B_W[i,j]$   \Comment{Step~1}
        \State initialize $dW[i,j]$ held by $B_W[i,j]$
        \For{$n = 0 \to N-1$}
            \State scatter $dY_n[i,j]$ to $B_A[i,j]$    \Comment{Step~2}
            \State all-gather $dY_n[i,j]$ within column \Comment{Step~3}
            \State $\tilde{dX}_n[:,j,i] = dY_n[:,j] \times W^T[j,i]$
            \State reduce-scatter $\tilde{dX}_n[:,j,i]$ within row  \Comment{Step~4}
            \State gather $dX_n[j,i]$ saved on $B_A[i,j]$   \Comment{Step~5}
            \State scatter $X_n^T[i,j]$ to $B_A[i,j]$       \Comment{Step~6}
            \State all-gather $X_n^T[i,j]$ within row       \Comment{Step~7}
            \State $dW[i,j] += X_n^T(i,:) \times dY_n[:,j]$
        \EndFor
        \State update $W[i,j]$ with $dW[i,j]$   \Comment{Step~8}
    \end{algorithmic}
    }
\end{algorithm}

\begin{figure*}[!t]
    \centering
    \includegraphics[width=\textwidth]{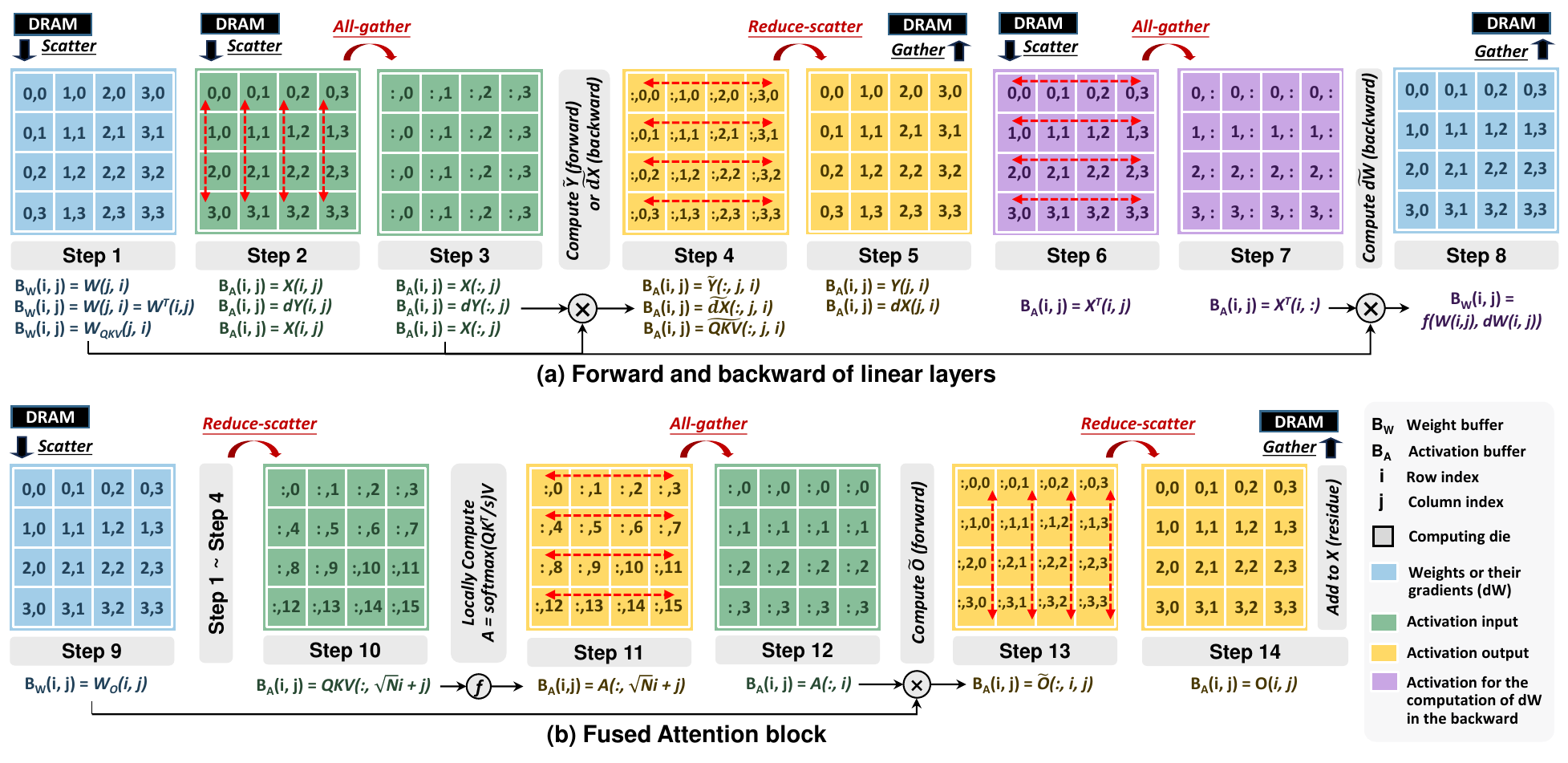}
    \caption{(a) shows the operations required to train a linear layer, corresponding to steps outlined in Algorithm~\ref{alg:main}. The formulae at the bottom summarize the hardware-to-software mapping. The three lines of equations below Steps~1-5 respectively correspond to (1) a general forward pass, (2) part of a general backward pass, and (3) part of an Attention block's forward. They share identical operations, differing only in variable. Steps~6-8 illustrate the computation of weight gradients, utilizing data prepared in Step 3. Figure (b) outlines operations for the forward of an Attention block, where multi-head attention~(Step~10-12) is inserted between two fused layers~(Step~1-4 and Step~9,13,14). 
    }
    \label{fig:method}
\end{figure*}

\label{chap:scheme}

The distributed training method transforms a single \textbf{global} data communication across all dies into two \textbf{local} communications within dies arranged in the same row or column. These two localized communication phases orchestrate with each other through the 2D mapping of the weight matrix. We carefully design the dataflow to ensure that the localized collective operations only include all-gather and reduce-scatter, which can fully utilize the D2D links on \sysname. Compared to the tensor parallelism used in Megatron, the volume of activation to be transferred is reduced through the co-design of 2D matrix tiling and communication scheme. Compared to Optimus, our collective operations are more efficient.

We introduce the detailed executions for FFN blocks and Attention blocks in Transformers as follows. It achieves reduced communication complexity, whose formulation will be shown Section~\ref{chap:theo}.

\subsection{FFN Blocks}
\label{chap:ffn}

The training process of a single linear layer is outlined in Algorithm~\ref{alg:main}, with crucial steps illustrated in Figure~\ref{fig:method}.
In each step, communication only happens among dies in the same column or row instead of all dies in the package, thus leading to less communication steps and data to transfer.
For software, $[i,j]$ represents the matrix tiling in the $i$th row and $j$th column, while for hardware, $[i,j]$ denotes the die's coordinates. Together, they describe the mapping from tensors to dies. Two details regarding the notation need attention. First, although the activation is three-dimensional $[b,s,h]$ (Section~\ref{chap:transformer}), during matrix multiplication, it can be treated as a two-dimensional tensor $[bs, h]$ without affecting the computation results. Second, the notation for partial sums (e.g., $\tilde{Y}$ and $\tilde{dX}$) is special, involving three elements. The first and last elements still represent the row and column indices, respectively, while the middle element indicates the input channel index of the weight. For example, in $Y=XW = \Sigma_{k=0} (X[:,k]W[k,:]) = \Sigma \tilde{Y}[:,k,:]$, $\tilde{Y}$ is the partial sum and need to be added together. 

One iteration comprises both forward and backward processes, with the latter being more complex due to the computation of $dW$. Our dataflow reuses the $dY$ that has already been all-gathered for the computation of $dX$ to reduce data transfer, as shown in Figure~\ref{fig:method}(a). The mapping of tensors is initially designated when tensors are fetched from DRAM, and NoP communication ensures the operands have been prepared in each die's local buffers before computation.

We make the following optimizations of NoP communication. First, the preparation of input activation~($X[:,j]$) is operated in two steps: scatter from DRAM (Step~2) and all-gather by NoP (Step~3). Compared to fetching X[:,j] from DRAM in a single step, this two-step operation substitutes repetitive expensive DRAM accesses with high-speed and low-energy D2D transfers, thus reducing communication overheads. Furthermore, in our method, NoP only performs two types of collective operations: all-gather and reduce-scatter, both of which can be efficiently executed on the ring topology mentioned in Figure~\ref{fig:architecture}(b).

The following paragraph describes the on-package execution when two linear layers are fused. Notably, at the end of the mini-batch's computation, the tiling of the activation~(Step~5) mirrors the transposition of that at the beginning~(Step~2). Consequently, the fused layer can be directly computed without additional communication, only requiring to transpose its weight when loading from DRAM~(Step~9). An FFN block comprises two linear layers. Therefore, after two rounds of transposition, the input and output mappings are identical, facilitating a direct residual link addition.

\subsection{Attention Blocks}

Attention Blocks differs from FFN blocks in the multi-head attention, whose executions are explained below. As Figure~\ref{fig:method}(b) shows, after computing the linear layers that generate $Q,K,V$~(Step~4), the output activation is partitioned along the hidden size dimension instead of the sequence length dimension as in FFN, and then reduce-scattered horizontally~(Step~10). This operation ensures that $Q, K, V$ from the same head are processed on the same die, thus utilizing the intrinsic parallelism provided by multiple heads and eliminating inter-die communication. If the number of dies surpasses the number of heads, activation of the same head is saved on different dies, and an all-reduce operation is necessary to compute the complete $A$. Afterwards, an all-gather operation converts the data layout as depicted in Step~12, facilitating the subsequent multiplication with $W_O$. Step~13 and Step~14 shows operations of the fused layer at the end of the Attention block.

In summary, attention blocks can be viewed as inserting multi-head attention between two fused linear layers with modification of collective operations at the merge points. The backward process of Attention blocks resembles that of FFN, with the key difference being that separate heads are handled by different dies, just like the forward process.

\section{Theoretical Analysis}
\label{chap:theo}

This Section provides the formalization of \sysname{}'s smaller communication overheads~(Section~\ref{chap:form}) and proves its weak scaling features~(Section~\ref{chap:weak_ana}).
For clarity in subsequent discussions, Table \ref{tab:para} lists the notations used and their descriptions.

\begin{table}[!t]
    \small
    \caption{Hardware parameters}
    \begin{tabular}{cl}
        \toprule
        Notation& Description\\
        \midrule
        $N$       & The number of computing dies on the package.\\
        $\alpha$  & D2D link latency. \\
        $\beta$   & D2D link bandwidth.\\
        $c$       & The number of DRAM channels. \\
        ${\beta}_{D}$ & DRAM bandwidth.\\
        \bottomrule
    \end{tabular}
    \label{tab:para}
\end{table}

\subsection{Formalization}
\label{chap:form}

We compare our method with the following three configurations: (1) Flat-ring, which employs 1D-TP and performs ring all-reduce, as used in Megatron~\cite{shoeybi2019megatron}; (2) Torus-ring, which also employs 1D-TP but performs 2D-torus all-reduce; and (3) Optimus, which employs 2D-TP and performs broadcast and reduce. ~\footnote{Given that existing works primarily target GPU systems, we make necessary adjustments to ensure they align with the hardware assumptions discussed in Section~\ref{chap:arch}}. 

\paragraph{Communication overheads}

\begin{table*}[!t]
  \centering
  \caption{Summary of NoP communication overheads when using different training methods. \sysname{} has a reduced complexity in both link latency and transmission time, thus being more scalable compared to the existing methods.}
  \footnotesize
  \label{tab:compare}
\begin{tabular}{ccccccccc}
    \toprule
    \multirow{2}{*}{Workload} & \multicolumn{4}{c}{\textbf{Link Latency}} & \multicolumn{4}{c}{\textbf{Transmission Time}} \\
    \cmidrule(lr){2-5} \cmidrule(lr){6-9}
    &  Flat-ring &  Torus-ring &  Optimus & \textbf{\sysname} & Flat-ring & Torus-ring & Optimus & \textbf{\sysname} \\
    \midrule
    Fwd Atten. & $2(N-1)\alpha$ & $4(N-\sqrt{N})\alpha$ & $4(N-\sqrt{N})\alpha$ & $8(\sqrt{N}-1)\alpha$ & $\frac{2(N-1)}{N}\gamma$ & $\frac{N-1}{N}\gamma$ & $\frac{\log_2 N}{2\sqrt{N}}(2\gamma+4\xi)$ & $\frac{6(\sqrt{N}-1)}{N}\gamma$ \\
    \midrule
    Fwd FFN & $2(N-1)\alpha$ & $4(N-\sqrt{N})\alpha$ & $4(N-\sqrt{N})\alpha$ & $8(\sqrt{N}-1)\alpha$ & $\frac{2(N-1)}{N}\gamma$ & $\frac{N-1}{N}\gamma$ & $\frac{\log_2 N}{2\sqrt{N}}(5\gamma+8\xi)$ & $\frac{10(\sqrt{N}-1)}{N}\gamma$ \\
    \midrule
    Bwd Atten. & $3(N-1)\alpha$ & $6(N-\sqrt{N})\alpha$ & $12(N-\sqrt{N})\alpha$ & $12(\sqrt{N}-1)\alpha$ & $\frac{3(N-1)}{N}\gamma$ & $\frac{3(N-1)}{2N}\gamma$ & $\frac{\log_2 N}{2\sqrt{N}}(4\gamma+8\xi)$ & $\frac{8(\sqrt{N}-1)}{N}\gamma$ \\
    \midrule
    Bwd FFN & $3(N-1)\alpha$ & $6(N-\sqrt{N})\alpha$ & $12(N-\sqrt{N})\alpha$ & $12(\sqrt{N}-1)\alpha$ & $\frac{3(N-1)}{N}\gamma$ & $\frac{3(N-1)}{2N}\gamma$ & $\frac{\log_2 N}{2\sqrt{N}}(10\gamma+16\xi)$ & $\frac{15(\sqrt{N}-1)}{N}\gamma$ \\
    \bottomrule
  \end{tabular}
\end{table*}

We decompose the NoP overheads into two parts: \textit{link latency} and \textit{transmission time}. The former is a fixed latency for each transmission due to the physical distance, while the latter is the time to transfer data chunks once the connection has been established, calculated as data size divided by D2D bandwidth. For simplicity, we assume that the number of dies $N$ is a perfect square.

We first derive the overheads for all-gather~(AG) and reduce-scatter~(RS) in our method, $L$ for link latency and $T$ for transmission time. Assuming total data volume is $S$, data with size of $\frac{S}{N}$ are communicated among {\fontsize{8pt}{10pt}\selectfont $\sqrt{N}$ } dies in the same row or column for {\fontsize{8pt}{10pt}\selectfont $\sqrt{N}$ }$-1$ steps, as illustrated in Figure~\ref{fig:all-reduce}(b). The latency of by-pass links is twice that of adjacent links, as mentioned in Section~\ref{chap:link}.
\begin{equation}
    \begin{aligned}
        L_{AG} = L_{RS} &= (\text{\footnotesize{$\sqrt{N}$}}-1)\times 2\alpha \\
        T_{AG} = T_{RS} &= (\text{\footnotesize{$\sqrt{N}$}}-1)\times \text{\footnotesize{$\frac{S}{N}$}} \times \text{\footnotesize{$\frac{1}{\beta}$}}
    \end{aligned}
\end{equation}
Data volume $S$ takes on different values when transferring different activations. Define $\gamma = bsh/ \beta$ and $\xi = h^2/ \beta$, with activations named as in Figure 2, then NoP overheads for the forward pass of Attention blocks and FFN blocks are:
\begin{equation}
    \begin{aligned}
        L_{\text{\textit{fwd\_Atten}}} = L_{\text{\textit{fwd\_FFN}}} = 2L_{AG} + 2L_{RS} = 8(\text{\footnotesize{$\sqrt{N}$}}-1)\alpha 
    \end{aligned}
\end{equation}
\begin{equation}
    \begin{aligned}
         T_{\text{\textit{fwd\_Atten}}} &= T_{AG\_X} + T_{RS\_QKV} + T_{AG\_A} + T_{RS\_O} \\
            & = (1+3+1+1)\text{\scriptsize{$\frac{\sqrt{N}-1}{N}$}}\gamma = 6\text{\scriptsize{$\frac{\sqrt{N}-1}{N}$}}\gamma
    \end{aligned}
\end{equation}
\begin{equation}
    \begin{aligned}
        T_{\text{\textit{fwd\_FFN}}} &= T_{AG\_O} + T_{RS\_Z} + T_{AG\_Z} + T_{RS\_X} \\
            & = (1+4+4+1)\text{\scriptsize{$\frac{\sqrt{N}-1}{N}$}}\gamma = 10\text{\scriptsize{$\frac{\sqrt{N}-1}{N}$}}\gamma
    \end{aligned}
\end{equation}

The backward process requires an additional all-gather of the original activations (Step~7), resulting in higher NoP overheads compared to the forward pass. For the Attention block, the extra transfers involve $X$ and $A$, while for the FFN block, they involve $X$ and $Z$. Due to space limitations, we omit the derivation since they are similar to the equation above. The results are summarized in Table~\ref{tab:compare}.

The drawbacks of existing methods can be summarized as follows. For 1D-TP, the communication volume is $\sqrt{N}$ times that of our method. Although 2D-torus provides more links than flat-ring, it can only reduce the constant factor of NoP overheads, not the complexity. For 2D-TP, the execution of broadcast and reduce operations is inefficient because they cannot utilize all available bandwidth. Their derivations can be found in original papers~\cite{xu2023efficient,ying2018image,tanakaimagenet}.

\paragraph{SRAM Capacity Requirements.} In our method, the maximum SRAM usage occurs when storing the all-gathered activation $Z$, whose size is $4sh/\sqrt{N}$. Therefore, increasing $N$ can relieve the memory burden. In contrast, 1D-TP requires storing complete activations such as $X$ and $O$ with size of $sh$ on every dies. As the model grows, they can exceed the fixed capacity of SRAM. Optimus needs extra storage for segments broadcast from other dies, further burdening the already capacity-constrained weight buffer.

\label{chap:layout}
\paragraph{Layout Constraints.} Our method does not impose specific constraints on the number and layout of dies. Conversely, the flat-ring necessitates an even number of dies to establish the Hamiltonian ring, while Optimus requires a square number of dies. Although the torus-ring method is not constrained by layout as well, it suffers severe performance degradation on rectangular dies due to the imbalanced transmission delay between shorter and longer links. 

\subsection{Weak Scaling Analysis} 
\label{chap:weak_ana}

Our method demonstrates efficient weak scaling, as demonstrated by maintaining nearly constant portions of computation, NoP communication, and DRAM access - the major components of system latency - when scaling both model dimensions and die counts. Specifically, when the model is scaled, the hidden dimension size $h$ increases by factor $k$, while the die count $N$ increases by $k^2$.

For computation latency $C(k)$, we can derive Equation~(\ref{equ:comp}). 
\begin{equation}
    C(k) = \frac{\text{computation}}{\text{computation power}}\propto \frac{h(k)^2}{N(k)} \propto \frac{k^2}{k^2} = 1
    \label{equ:comp}
\end{equation}

For NoP latency $T(k)$, we have Equation~\ref{equ:NoP}. The analysis focuses solely on the transmission time and omitting the link latency, as the latency coefficient $\alpha$ is significantly smaller than the bandwidth coefficient $\gamma$, which will be demonstrated in Section ~\ref{chap:exp_link}.
\begin{equation}
    T(k) \propto \text{\scriptsize{$\frac{\sqrt{N(k)}-1}{N(k)}$}}\gamma \sim \text{\small{$\frac{1}{\sqrt{N(k)}}$}}h(k) \propto \frac{k}{\sqrt{k^2}} = 1
    \label{equ:NoP}
\end{equation}

For DRAM access $D(k)$, we have Equation~\ref{equ:memory}. We focus on DRAM access for activation and overlook that for weights, because weights are reused across multiple batches, rendering it a minor contributor to system latency, as demonstrated in Section~\ref{chap:exp_weak}. The number of DRAM channels $c$ grows with the package's perimeter as explained in Section~\ref{chap:arch_dram}.
\begin{equation}
    D(k) \propto \frac{\text{activation volume}}{\text{DRAM bandwidth}} \propto \frac{h(k)}{c(k)\beta_{DRAM}} \propto \frac{k}{k} = 1
    \label{equ:memory}
\end{equation}

We ensure that the SRAM capacity requirements also remain constant, thus our method will always be valid regardless of the model size. We use $U_W(k)$ and $U_A(k)$ representing the maximal usage of weight buffer and activation buffer.
\begin{equation}
    \begin{aligned}
        &U_W(k) \propto \frac{\text{weight volume}}{\text{number of dies}} \propto \frac{h(k)^2}{N(k)} \propto \frac{k^2}{k^2} = 1\\
        &U_A(k) \propto \text{all-gathered act. volume} \propto \frac{h(k)}{\sqrt{N(k)}} \propto \frac{k}{k} = 1
    \end{aligned}
\end{equation}

The enhanced transmission performance of our approach stems from the 2D layout and connection of computing dies. Theoretically, this scheme could extend to GPU-based distributed systems. However, migrating this scheme from Chiplet to GPU clusters would lead to significant performance degradation due to the differences in latency and bandwidth of inter-GPU and inter-server links. Faster links would wait for slower ones, thus harming the utilization of inter connections.

\section{Evaluation}
\label{chap:eval}

\subsection{Experiment Setting}
\label{chap:setting}

\begin{figure*}[h]
    \centering
    \includegraphics[width=\textwidth]{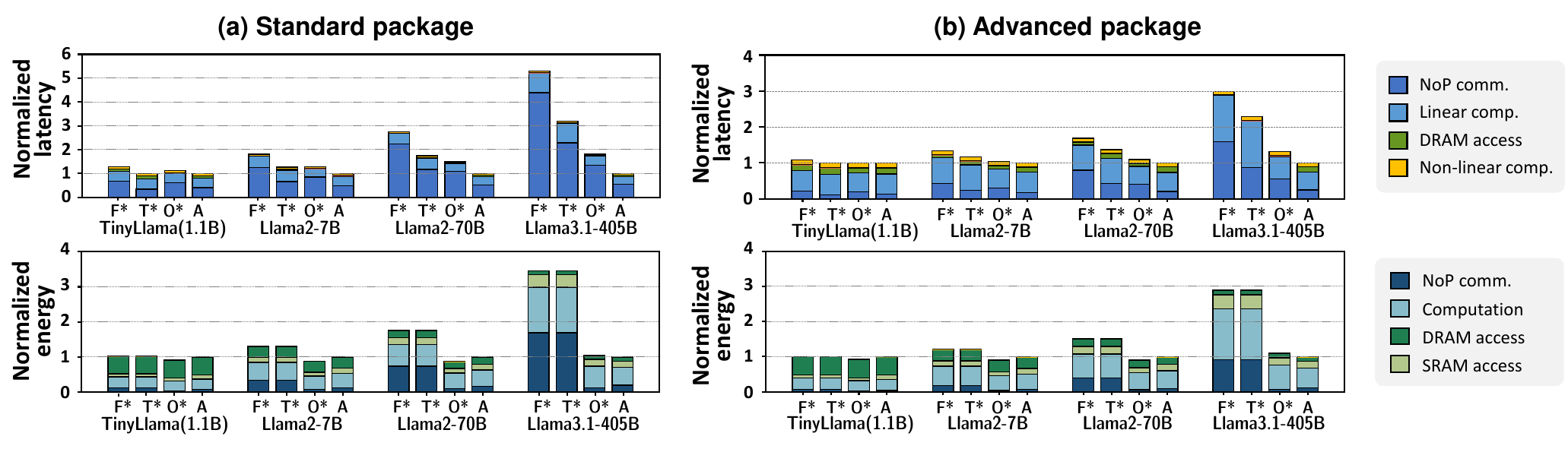}
    \caption{Comparison of \sysname{} with other distributed methods under various workloads and different forms of package. F, T, O, A respectively stand for (1) 1D-TP with ring all-reduce which is used by Megatron~\cite{shoeybi2019megatron}, (2) 1D-TP with 2D-torus all-reduce, (3) Optimus~\cite{xu2023efficient}, and (4) \sysname{}, aligning with theoretical analysis in Section~\ref{chap:form}. Both latency and energy for each workload are normalized to \sysname{}. Methods marked with an asterisk (*) are practically invalid since they need SRAM capacity exceeding the hardware settings of 8MB weight/activation buffers. Since we employ the latency hiding technique as mentioned in \ref{chap:arch_dram}, the latency breakdown of DRAM access denotes the segment exceeds the on-package execution, rather than the entire DRAM access time. \sysname{} exhibits obvious improvements for both latency and energy, especially on larger workloads.
    }
    \label{fig:performance}
\end{figure*}

\paragraph{\textbf{Hardware configurations.}}
The experiments are conducted on simulated chiplet architectures as described in Figure~\ref{fig:architecture}. Most digital functional modules within the computing die are realized in RTL and synthesized using Synopsys Design Compiler~\cite{synopsys}, employing 28nm CMOS technology and achieving clock frequencies of up to 800MHz. Energy consumption for each module was estimated with PrimeTimePX. The area and read/write energy of SRAM buffers are derived from SRAM Compiler~\cite{compiler}. Based on TSMC reports~\cite{wu201316nm, wu20167nm}, we then rescale the area and power to 7nm, which has been adopted by \sota{} waferscale engines~\cite{lie2022cerebras} and GPUs~\cite{A100}. The computing die has an estimated area of 30.08$mm^2$. Each computing die comprises a 4$\times$4 PE array with 32 lanes per PE, complemented by an 8MB activation buffer and an 8MB weight buffer. Although fine-grained mapping and data reuse are not the focus of this paper, we model and validate them against the accelerator simulator Timeloop~\cite{parashar2019timeloop} to ensure accurate results. Our performance model yields consistent utilization and SRAM reuse results with Timeloop.

D2D link parameters including the link latency, bandwidth, energy and area are sourced from the Universal Chiplet Interconnect Express (UCIe) standards and supplemented by prior research~\cite{yin2023aries, waferscaleGPU}. While both advanced and standard packages transfer at a 16GT/s data rate, the advanced package's finer pitch enables higher link density, resulting in a \textit{higher bandwidth} within the same area constraint. For memory, we use DDR5-6400 as the DRAM, with latency aligned to stream trace simulations by Ramulator2~\cite{luo2023ramulator}. The bandwidth and energy consumption are set 51.2GB/s and 19pJ/bit respectively based on~\cite{DDR5, o2017fine}.

\paragraph{\textbf{Workloads.}} We evaluate Llama models with successively doubled hidden sizes ($h$): TinyLlama-1.1B ($h$=2048)~\cite{zhang2024tinyllama}, Llama2-7B ($h$=4096)~\cite{touvron2023llama2}, Llama2-70B ($h$=8192)~\cite{touvron2023llama2}, and Llama3.1-405B~($h$= 16384)~\cite{dubey2024llama3}. Their training systems scale proportionally, integrating 16, 64, 256, 1024 computing dies respectively. 
The batch size $b$=1024. We set the sequence length $s$=2048 when experiments involve TinyLlama to accommodate its positional embedding limits. Otherwise, we use each model's original pretraining sequence length\footnote{llama3.1-405B supports sequence length up to 128k, which is achieved through a standard pre-training and a subsequent training increasing the context window. Here we only consider the standard pre-training.}. 
Other parameters (GQA and intermediate dimensions) are obtained from the models' Huggingface.

\subsection{Overall Comparison}

Figure~\ref{fig:performance} demonstrates that \sysname{} has obvious advantages in latency and energy efficiency over other methods across different workloads regardless of the package type. 

For latency, we achieve at most $5.29\times$ and $3.00\times$ speedup when adopting standard package and advanced package, respectively, compared to the tensor parallelism used in Megatron~\cite{shoeybi2019megatron}. On larger workloads equipped with more computing dies, \sysname{}'s improvement is more obvious, which stems from the reduced NoP overheads. 

DRAM access only accounts for a small portion, as weight access is amortized across multiple batches, while activation access is overlapped by on-package communication and computation. In scaled systems, 1D-TP based methods exhibit increased computation time despite unchanged theoretical FLOPS per die, primarily due to the reduced PE array utilization. In contrast, 2D-TP methods maintain a more stable computation time through matrix tiling with balanced input and output channel counts. Regarding energy, our approach achieves improvements of up to $3.46\times$ and $2.89\times$ compared to Megatron TP for different packages. Unlike latency, where NoP communication constitutes a significant portion, energy consumption is primarily determined by computation. Both Optimus and \sysname{} demonstrate significantly reduced NoP overhead due to reduced data transfer volume.

SRAM overflow occurs for all methods except \sysname. This is because their peak SRAM requirements increase with the system scale as analyzed in Section~\ref{chap:form}, eventually exceeding the fixed SRAM buffer capacity. In contrast, \sysname{} maintains roughly constant SRAM requirements, ensuring feasibility across problem sizes.

\label{chap:exp_weak}

\subsection{Scaling Performance}

\begin{figure}[t!]
  \centering
  \includegraphics[width=\linewidth]{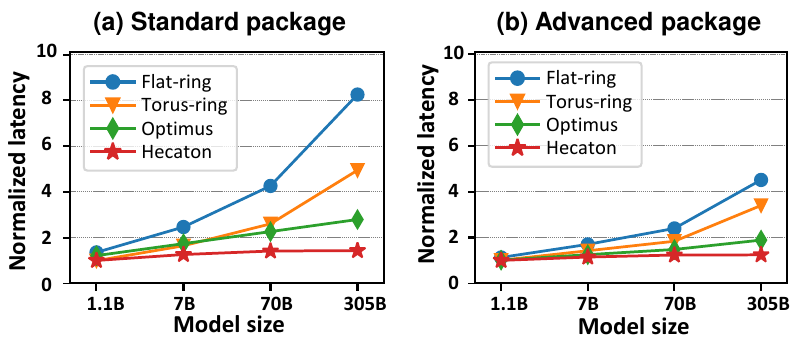}
  \caption{Study of scalability. The latency is normalized to the processing time for the smallest model. \sysname{} maintains roughly constant processing time as predicted by theoretical analysis in Section~\ref{chap:weak_ana}, while others cannot.}
  \label{fig:weak}
\end{figure}

Figure~\ref{fig:weak} shows the training latency of models with varied scales using different methods. The results verify theoretical analysis in Section~\ref{chap:weak_ana} that in \sysname{}, processing time remains approximately constant as the problem scales. In contrast, existing methods exhibit increasing latency since their NoP communication complexity surpasses other component complexities, which ultimately become the system bottleneck when the number of dies increases. This effect is more obvious when adopting standard packaging, whose lower D2D bandwidth results in proportionally higher NoP overhead in the system, and the gap between different methods becomes pronounced.

\subsection{The Impact of DRAM bandwidth}
\label{exp:memory}

We analyze the impact of DRAM bandwidth on system latency using three memory configurations: DDR4-3200 (previous generation), DDR5-6400 (current generation), and HBM2 (high-cost, high-end applications). There are two main observations. First, performance improvements brought by higher memory bandwidth saturates. Once the latency of DRAM access matches the latency of on-package execution, further increasing bandwidth only yields limited gains since computation and NoP communication have become the primary bottlenecks. Second, systems utilizing advanced packaging are more sensitive to the DRAM bandwidth, as the reduced NoP latency hides less DRAM access latency. This experiment demonstrates that common DDR already provides sufficient performance for our training system.

\begin{figure}[t!]
  \centering
  \includegraphics[width=\linewidth]{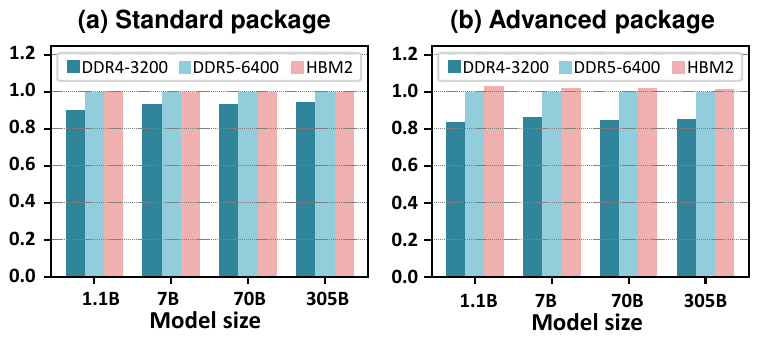}
  \caption{The impact of DRAM bandwidth on system latency. Speedup is normalized to DDR5-6400.}
  \label{fig:dram_latency}
\end{figure}

\subsection{The Impact of Link Latency}
\label{chap:exp_link}

Table~\ref{tab:latency} demonstrates link latency's proportion in system latency when $\alpha$=10ns, where adaptive and physical layers take 2ns as specified in UCIe. While its proportion increases as the system scales or adopts advanced packaging with higher bandwidth, the link latency's contribution to overall system performance remains small. This justifies our decision to omit link latency in the theoretical analysis in Section~\ref{chap:weak_ana}.

\begin{table}[t!]
    \caption{The proportion of link latency in system latency.}
    \vspace{-8pt}
    \label{tab:latency}
    \small
    \begin{center}
    \renewcommand{\arraystretch}{1.2}
        \begin{tabular}{c||c|c|c|c}
            \hline
            \textbf{Workload} & \textbf{llama-1.1B} & \textbf{llama-7B} & \textbf{llama-70B} & \textbf{llama-405B} \\
            \hline
            \textbf{Standard} & 0.549\% & 1.073\% & 2.127\% & 4.399\% \\
            \hline
            \textbf{Advanced} & 0.832\% & 1.787\% & 3.687\% & 7.678\% \\ 
            \hline
        \end{tabular}
    \end{center}
\end{table}
\vspace{-3pt}

\subsection{The Impact of Layout}
\label{chap:exp_layout}

\begin{figure}[t]
  \centering
  \includegraphics[width=\linewidth]{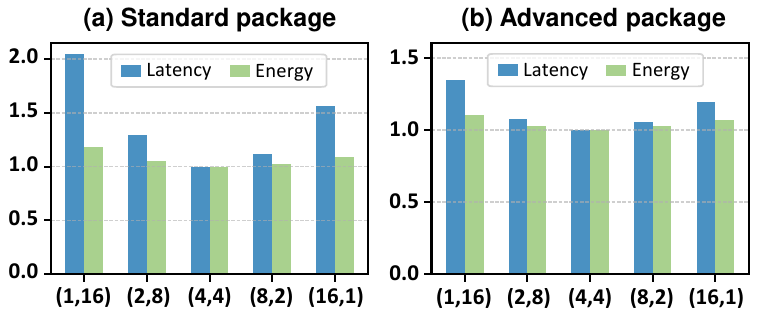}
  \caption{The impact of computing dies' layout on system latency / energy. Altogether 16 dies are arranged in (length, width). All metrics are normalized to the sqaure layout.}
  \label{fig:layout}
\end{figure}

\sysname{} can accommodate various layouts, and obtains the best latency and energy when dies are arranged in square, as depicted in Figure \ref{fig:layout}. For rectangular layout, it has a preference for longer width, which arises from the asymmetry in the weight matrix's dimensions. For example, when performing the scale-up linear layer in FFN blocks, both input and output activation necessitate communication, but the latter has a larger size. Matching the larger activation to a short side leads to transferring large data chunks in fewer communication steps, thus lifting the overall performance.

\subsection{Comparison with GPUs}

We evaluate the energy efficiency (FLOPS/W) of \sysname{} against GPU clusters for training Llama2-70B, which was originally trained on 7nm A100 GPUs. We compute the energy efficiency of distributed GPU system using GPU hours and power consumption reported in \cite{touvron2023llama2}. Compared to that, \sysname{} achieves a \textbf{22.36$\times$} improvement. This enhancement stems from three key factors. For computation, \sysname{} uses ASIC computing dies designed for large matrix multiplication, while general purpose GPU contains other components like cuda cores besides tensor cores. For communication, \sysname{} utilizes more compact on-package interconnection with higher bandwidth, lower latency, and reduced energy per bit. Our distributed training method further decreases the NoP overheads. For memory, \sysname{} leverages distributed SRAMs to enlarge on-chip storage, thereby eliminating repetitive DRAM accesses.

\section{Related Work}

\textbf{Chiplet-based architecture. }Advancements in packaging technology~\cite{mahajan2016embedded,soic,sow,huang2021wafer,lall1993overview} have excited interest in chiplet architectures. Advanced waferscale chiplets can occupy areas of several tens of thousands of $mm^2$ and integrate more than 1,000 dies~\cite{2048chiplet, waferscaleGPU, feng2024switch}. Chiplets' scaled-out computational power has been leveraged for DNN, pioneered by Simba~\cite{zimmer20190,zimmer20200,shao2019simba} and further optimized by subsequent works~\cite{tan2021nn,cai2024gemini,hao2023monad,deepburningseg,pal2020design,cai2024abss,xie2022transferable} through modeling and exploration of larger design spaces. With the rise of LLMs, several recent works~\cite{peng2023chipletcloud, CambriconLLM} have extended chiplets to LLM inference, primarily addressing the memory bottleneck through on-chip caching and advanced packaging. However, LLM training remains unexplored, where existing solution assumptions are invalidated by more severe memory and communication challenges, and the backward propagation introduces more complex dataflow.

\textbf{LLM training and finetuning.} ZeRO~\cite{rajbhandari2020zero,rajbhandari2021zero,ren2021zero} 
series of works perform data parallelism and reduce memory usage by dividing weights and gradients into shards or offloading them to CPU memory. Many works study~\cite{huang2019gpipe,li2021chimera, narayanan2019pipedream, narayanan2021memory,li2021terapipe} pipeline parallelism with smaller cold-start bubbles to lift the GPU utilization. These parallelisms are orthogonal to our method and can be utilized together to accelerate the LLM training. Megatron~\cite{shoeybi2019megatron} proposed 1D-TP tensor parallelism requiring all-reduce operations. Subsequent works introduce 2D~\cite{xu2023efficient} and 2.5D~\cite{wang2022tesseract} tensor parallelism, which theoretically reduce the complexity of data transfer but impose new requirements on topology. Compared with them, our method has a lower asymptotic communication complexity and fully utilizes the new architectural opportunity brought by chiplet.

\section{Conclusion}
In this paper, we propose \sysname, a scalable and cost-effective Chiplet architecture for LLM training and finetuning with scheduling minimizing DRAM access impacts. Along with the novel distributed method that co-designs matrix tiling and communication schemes, the system can achieve weak scaling and obtain obvious improvements in both latency and energy compared to previous works.

\bibliographystyle{plain}
\bibliography{reference}

\end{document}